\documentclass[aps,pre,amsmath,amssymb,reprint,groupedaddress,superscriptaddress]{revtex4-1}

\usepackage{graphicx}% Include figure files
\usepackage{dcolumn}% Align table columns on decimal point
\usepackage[colorlinks,linkcolor=blue,breaklinks]{hyperref}
\usepackage{chngcntr}
\usepackage[utf8]{inputenc}
\usepackage[T1]{fontenc}
\usepackage{newtxtext,newtxmath}

\newcommand{\angstrom}{\mbox{\normalfont\AA}}

\begin{document}

\title{Stability of Pinned Surface Nanobubbles Against Expansion: Insights from Theory and Simulation}

\author{Yawei Liu}
\email[]{yawei.liu@sydney.edu.au}
\affiliation{ARC Centre of Excellence in Exciton Science, School of Chemistry, University of Sydney, Sydney, New South Wales 2006, Australia}

\author{Stefano Bernardi}
\email[]{stefano.bernardi@sydney.edu.au}
\affiliation{ARC Centre of Excellence in Exciton Science, School of Chemistry, University of Sydney, Sydney, New South Wales 2006, Australia}

\author{Asaph Widmer-Cooper}
\email[]{asaph.widmer-cooper@sydney.edu.au}
\affiliation{ARC Centre of Excellence in Exciton Science, School of Chemistry, University of Sydney, Sydney, New South Wales 2006, Australia}
\affiliation{The University of Sydney Nano Institute, University of Sydney, Sydney, New South Wales 2006, Australia}

%\date{\today}

\begin{abstract}
While growth and dissolution  of surface nanobubbles has been widely studied in recent years, their stability under pressure changes or a temperature increase has not received the same level of scrutiny. Here, we present theoretical predictions based on classical theory for pressure and temperature thresholds ($p_c$ and $T_c$) at which unstable growth occurs for the case of air nanobubbles on a solid surface in water. We show that bubbles subjected to pinning have much lower $p_c$ and higher $T_c$ compared to both unpinned and bulk bubbles of similar size, indicating that pinned bubbles can withstand a larger tensile stress (negative pressure) and higher temperatures. The values of $p_c$ and $T_c$ obtained from many-body dissipative particle dynamics (MDPD) simulations of quasi-two-dimensional (quasi-2D) surface nanobubbles are consistent with the theoretical predictions, provided that the lateral expansion during growth is taken into account. This suggests that the modified classical thermodynamic description is valid for pinned bubbles as small as several nanometers. While some discrepancies still exist between our theoretical results and previous experiments, further experimental data is needed before a comprehensive understanding of the stability of surface nanobubbles can be achieved.
\end{abstract}
 
\maketitle

\section{Introduction}
The stability of surface nanobubbles has been a controversial topic since the work of Parker \textit{et al.} in 1994~\cite{Parker1994}, but it experienced renewed interest after the first AFM images were obtained in 2000~\cite{Lou2000,Ishida2000a}. Nanoscale bubbles at solid-water interfaces are predicted to be unstable due to an extremely high inner pressure. For example, according to the Epstein-Plesset theory~\cite{Epstein1950}, the lifetime of bubbles smaller than $1000$ nm in water should be less than $0.02$ s, which would also make their experimental detection extremely difficult~\cite{Lohse2015a,Alheshibri2016}. However, many experimental techniques including AFM~\cite{Lou2000,Ishida2000a,Zhang2006b}, rapid cryofixation~\cite{Switkes2004}, neutron reflectometry~\cite{Steitz2003}, and direct optical visualization~\cite{Karpitschka2012,Chan2012} have confirmed that surface nanobubbles can exist for hours or even days.

In 2013, experimental, theoretical, and computational studies proved the contact line pinning effect to be crucial for the stability of surface nanobubbles~\cite{Zhang2013,Weijs2013,Liu2013d}. 
Pinned bubbles, whose contact line is anchored by surface heterogeneities, can reach thermodynamically stable states under oversaturation, or dissolve on a much longer timescale than bulk bubbles when undersaturated. 
Since 2013, an impressive number of studies of surface nanobubbles have been performed to investigate the contact line pinning effect~\cite{Lohse2015b,Chan2015,Guo2016,Dollet2016,Maheshwari2016,Xiao2017,Tan2017,Tan2018,Maheshwari2018}. These studies examined surface nanobubbles under pressures and temperatures at which they were expected to dissolve due to gas diffusion, but were instead found to be stable. Two experimental papers studied the evolution of surface nanobubbles under pressure reduction (i.e., in the cavitation regime) and temperature increase. At a specific pressure or temperature threshold ($p_c$ and $T_c$), the bubble undergoes expansion and suffers an unstable and uncontrolled growth. As early as 2007, Borkent \textit{et al.} reported that surface nanobubbles do not cavitate down to $-6$ Mpa~\cite{Borkent2007}, a much lower value than the Blake threshold below which a bulk bubble with the same radius of curvature starts to grow unbounded~\cite{Blake1949,Brennen1995}. In 2014, Zhang \textit{et al.} unexpectedly observed that surface nanobubbles also remain stable when the temperature increases up to a value close to the boiling point of water~\cite{Zhang2014a}. However, despite the surprising stability that surface nanobubbles have shown against expansion induced by both pressure reduction and temperature increase (referred to as \textit{superstability} in Ref.~\cite{Borkent2007,Zhang2014a}), only a few studies have been dedicated to this phenomenon so far. One reason for this may be the lack of experimental techniques capable of probing surface nanobubbles without perturbing them.

More recently, using molecular dynamics simulations, Dockar \textit{et al.}~\cite{Dockar2019} studied the mechanical stability of surface nanobubbles pinned on patterned substrates and exposed to a pressure reduction. They found a much lower $p_c$ than that predicted by the Blake threshold. Their simulation results agreed with a corrected cavitation threshold they derived for pinned bubbles. In this work, we investigate the stability of pinned surface nanobubbles under pressure reduction as well as temperature increase by extending the classical theory to pinned bubbles and using particle dynamics simulations. We first present a theoretical analysis of $p_c$ and $T_c$ for both unpinned and pinned surface nanobubbles. The stability of surface nanobubbles in water is discussed in detail based on the theoretical calculations. We then present results for $p_c$ and $T_c$ obtained from many-body dissipative particle dynamics (MDPD) simulations of quasi-two-dimensional (quasi-2D) pinned surface nanobubbles and compare them with the corresponding theoretical predictions. Finally, we compare our results with earlier studies, including experimental work.

\section{Theory}

\begin{figure}[tb]
	\begin{center}
		\includegraphics[width=0.4\textwidth]{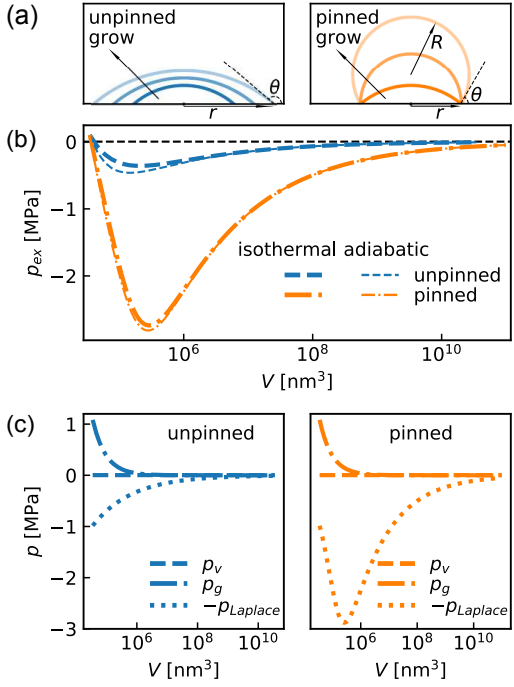}
		\caption{(a) Schematic illustrating the difference in growth mechanism between unpinned and pinned surface nanobubbles. (b) The change in the excess pressure $p_{ex}$ (see Eq.~\ref{eq:pex}) for growing surface nanobubbles with $r_0=50$ nm and $\theta_0=160^{\circ}$ at room temperature and pressure (i.e., $T = T_0=298$ K and $p_{atm}=101.325$ kPa). Results are shown for both isothermal and adiabatic conditions. The horizontal dashed line corresponds to $p_{v,sat}=3.158$ kPa at $T=298$ K. (c) The change in the three contributions to $p_{ex}$, i.e. $p_{v}$, $p_{g}$ and $-p_{Laplace}$, for unpinned and pinned surface nanobubbles under isothermal conditions.}
		\label{fig:fig01}
	\end{center}
\end{figure}

We first consider an equilibrium surface nanobubble in a liquid with a lateral radius $r$ and a liquid-side contact angle $\theta$ [see Fig.~\ref{fig:fig01} (a)]. The bubble is a spherical cap, with a radius of curvature $R=r/\sin \theta$, and volume $V=\pi r^3 g(\theta)/3\sin^3\theta$ with $g(\theta)=(2-\cos\theta)(1+\cos \theta)^2$. The evolution of an isolated spherical bubble is described by the Rayleigh-Plesset equation~\cite{Rayleigh1917,Plesset1949,Plesset1977},
\begin{equation}\label{eq:RP_eq}
R\ddot{R}+\frac{3}{2}\dot{R}^2+4v\frac{\dot{R}}{R} = \frac{p_b(t)-p_l(t)-2\gamma/R}{\rho_l}.
\end{equation}
Here, $t$ is the time, $\rho_l$ is the liquid density, $\gamma$ is the surface tension, $v$ is the kinematic viscosity of the liquid, $p_l(t)$ is the pressure in the liquid far away from the bubble surface, and $p_b(t)$ is the pressure inside the bubble. The right-hand side in Eq.~\ref{eq:RP_eq} contains the "forces" driving the bubble to grow or shrink: the collapsing force $F_c=p_l+2\gamma/R$ , which is the sum of the liquid pressure and the Laplace pressure; and the expanding force $F_e=p_b=p_g+p_v$, given by the sum of the gas pressure ($p_g$) and the vapor pressure ($p_v$) inside the bubble. In general, the bubble tends to grow if $F_e>F_c$, shrink when $F_e<F_c$, or remain in equilibrium when $F_c=F_e$. A bubble on a solid surface is expected to undergo a similar evolution to that of a spherical bubble in the bulk~\cite{Bremond2006a}. Therefore, assuming that the bubble responds quasi-statically to an external perturbation, we can use the same expansion-collapse criterion to predict its fate under a pressure or temperature change.

The surface nanobubble is in equilibrium at $t=0$ with $F_c=F_e$, i.e., $p_{l0}+2\gamma (T_0)/R_0=p_v(T_0)+p_{g0}$ where $T$ is the temperature and subscript $0$ denotes the initial conditions at $t=0$. If $p_{l0}$ is equal to the atmospheric pressure ($p_{atm}$), we have
\begin{equation}\label{eq:pg0eq}
p_{g0}=p_{atm}+ \frac{2\gamma(T_0) \sin\theta_0} {r_0} - p_v(T_0,r_0,\theta_0).
\end{equation}
The vapor pressure $p_v$ can be estimated via the Kelvin equation, i.e., $\ln\frac{p_v(T,r,\theta) }{p_{v,sat}(T)} = \frac{-2\gamma(T) V_m(T)\sin\theta}{r \bar{R} T}$ where $p_{v,sat}$ is the saturated vapor pressure for a flat liquid-vapor interface, $\bar{R}$ the gas constant, and $V_m$ the molar volume of liquid defined as the molar mass ($M$) divided by its density, i.e., $V_m(T)=M/\rho_l(T)$.

\subsection{No gas diffusion}
If the pressure/temperature perturbation takes place too rapidly for significant gas diffusion, there is no appreciable mass transfer of gas to or from the liquid. The evolution of the gas pressure can then be derived in two limiting cases of heat transfer between the liquid and the bubble: 1) isothermal evolution, i.e. the bubble has the same temperature as the liquid, in which case $p_g = p_{g0} V_0 T / V T_0$ for an ideal gas; 2) adiabatic evolution, in which $p_g = p_{g0} \left( V_0 / V \right)^k$ with $k$ the adiabatic index. Considering that unstable growth occurs when $F_e>F_c$ during the whole growth process, this gives us
\begin{multline}\label{eq:pex}
p_l < p_{ex}(r,\theta,T) = p_v(T,r,\theta) - \frac{2\gamma(T) \sin\theta}{r}  \\ + p_{g0} \left( \frac{3V_0\sin^3\theta}{\pi r^3 g(\theta)} \right)^n \left(\frac{T}{T_0}\right)^m,
\end{multline}
with $m=n=1$ for the isothermal evolution and $m=0$, $n=k$ for the adiabatic evolution. Here $p_{ex}(r,\theta,T)$ denotes the excess pressure against the liquid pressure. We also assume that the vapor pressure and the surface tension only depend on the liquid temperature which is always uniform.

\begin{figure*}[tb]
	\begin{center}
	    \includegraphics[width=0.8\textwidth]{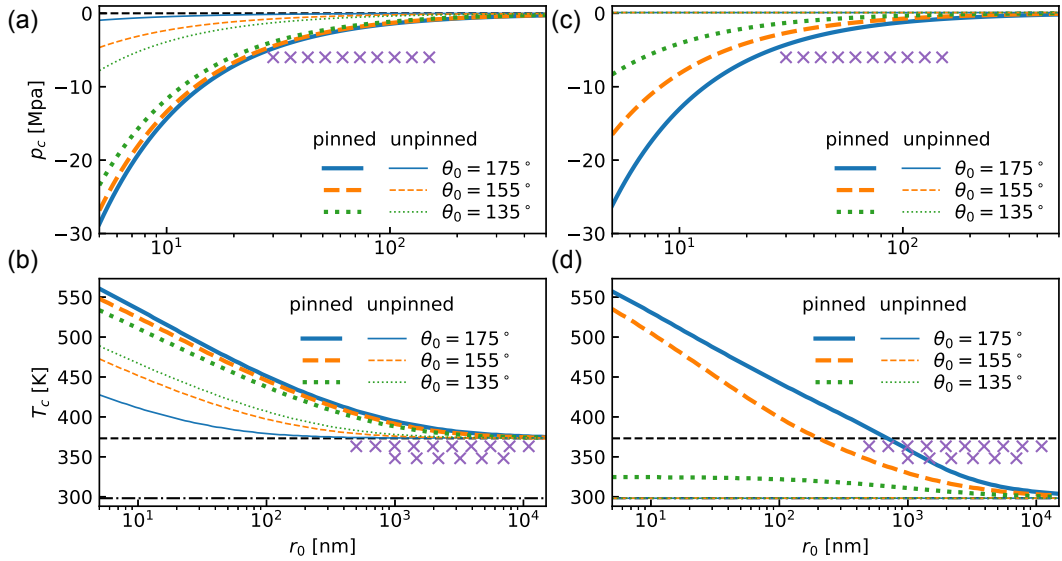}
		\caption{$p_c$ and $T_c$ in Eq.~\ref{eq:pex} and Eq.~\ref{eq:pex_henry} as a function of $r_0$ when $\theta_0=135^\circ$, $155^\circ$ and $175^\circ$ for surface nanobubbles in water initially at room temperature and pressure. The purple crosses indicate surface nanobubbles that have been reported to be experimentally stable~\cite{Borkent2007,Zhang2014a}. (a-b) $p_c$ and $T_c$ with no gas diffusion across the bubble surface and isothermal evolution (i.e., from Eq.~\ref{eq:pex} with $m=n=1$). The horizontal dashed lines correspond to $p_{v,sat}=3.158$ kPa in (a), and the water boiling point $373$ K and the room temperature $298$ K in (b). (c-d) $p_c$ and $T_c$ with gas equilibration (i.e., from Eq.~\ref{eq:pex_henry}). The horizontal dashed line in (d) corresponds to the boiling point $373$ K.}
		\label{fig:fig02}
	\end{center}
\end{figure*}

Equation~\ref{eq:pex} describes the \textit{ unstable growth condition} for surface nanobubbles: the excess pressure ($p_{ex}$) has to be always larger than the liquid pressure (${p_l}$). The change in $p_{ex}(r,\theta,T)$ depends on how the bubble evolves on the substrate. Ideally, we can identify two simplified scenarios~\cite{Liu2013c,Snoeijer2013}: a constant contact angle model (i.e., $\theta\equiv \theta_0$) for unpinned bubbles and a constant lateral radius model (i.e., $r\equiv r_0$) for pinned bubbles [see Fig.~\ref{fig:fig01} (a)]. We note that the unstable growth condition for unpinned bubbles is equivalent to that of bulk spherical bubbles with the same $R_0$, and that Eq.~\ref{eq:pex} leads to the Blake threshold equation for bulk bubbles exposed to a pressure reduction~\cite{Blake1949,Brennen1995} (see Appendix~\ref{appendixA}).

As an example, Figure~\ref{fig:fig01} (b) shows how the $p_{ex}(r,\theta,T)$ given by Eq.~\ref{eq:pex} changes for surface nanobubbles growing in water at room temperature and pressure (i.e., $T = T_0=298$ K and $p_{atm}=101.325$ kPa) when $r_0=50$ nm and $\theta_0=160^{\circ}$. The vapor pressure, the liquid density and the surface tension of water were calculated using the empirical equations in Ref.~\cite{Vargaftik1983,Wagner1993} (see Appendix~\ref{appendixB}), and $k=1.4$ for the adiabatic evolution. Initially, $p_{ex}=p_l=p_{atm}$, but as the bubble grows $p_{ex}$ decreases to a minimum value, then increases and gradually approaches $p_{v,sat}$. Pinned bubbles, however, have a much lower minimum $p_{ex}$ than unpinned ones, i.e. $-0.36$ Mpa and $-0.46$ Mpa for the former, and $-2.74$ MPa and $-2.81$ MPa for the latter. In both cases, a slightly lower minimum $p_{ex}$ is observed for adiabatic vs isothermal conditions.

These results can be understood by considering the three terms that contribute to $p_{ex}$: the vapor pressure $p_v$, the gas pressure $p_g$, and the Laplace term $-p_{Laplace}=-2\gamma \sin \theta /r = -2\gamma / R$. These are plotted independently in Fig.~\ref{fig:fig01} (c) for the isothermal case. For both unpinned and pinned bubbles, $p_v$ remains essentially constant and negligible at room temperature while $p_g$ decreases from $p_{g0}$ as the bubble expands. The change in the Laplace pressure, however, differs qualitatively depending on whether the bubble is pinned or not. For the unpinned bubble, $-p_{Laplace}$ keeps increasing because $R$ always increases during the growth [see the left image in Fig.~\ref{fig:fig01} (a)], which means that $p_{ex}$ only exhibits a minimum because $-p_{Laplace}\propto ~1/r$ decays slower than $p_g\propto 1/r^3$ (see Eq.~\ref{eq:pex}). In contrast, for the pinned bubble, $-p_{Laplace}$ decreases while $\theta>90^\circ$, reaches a minimum at $\theta=90^\circ$ when $R=r_0$, and then increases for $\theta<90^\circ$ [see the right image in Fig.~\ref{fig:fig01} (a)]. As a result, the $p_{ex}$ is much larger in magnitude at the critical state. Due to the influence of $p_g$, $\theta$ at the critical state is smaller than $90^\circ$ (\textit{e.g.} $\theta\sim 86^\circ$ at the critical state for the pinned bubble above). In comparison, the adiabatic evolution results in a slightly more negative $p_{ex}$ because $p_g\propto 1/r^{3k}$ decays faster [see Fig.~\ref{fig:fig01} (b)].

To induce unstable growth, one can decrease the liquid pressure until $p_l$ is below a critical pressure $p_c$ equal to the minimum $p_{ex}$. Our analysis shows that pinned bubbles have much lower $p_c$ than unpinned ones. Alternatively, one can increase the liquid temperature until the minimum $p_{ex}$ is larger than the ambient pressure $p_l=p_{atm}$. Thus, there exists a critical temperature $T_c$ above which unstable growth occurs. Pinned bubbles have higher $T_c$ as $p_{ex}$ is lower. For example, for the surface nanobubbles considered above, $T_c\sim 423$ K if unpinned, while $T_c\sim 473$ K if pinned (see Appendix~\ref{appendixC}). Therefore, pinned surface nanobubbles are much more stable than unpinned ones under either a pressure reduction or a temperature increase. %Here, $p_c$ and $T_c$ are also the pressure and temperature thresholds beyond which unstable growth occurs during the pressure/temperature changes.

Figure~\ref{fig:fig02} (a) and (b) give the results for $p_c$ and $T_c$ as a function of $r_0$ at several values of $\theta_0$ for surface nanobubbles initially at room temperature and pressure and then undergoing isothermal evolution, obtained by numerically solving Eq.~\ref{eq:pex}. For all bubbles, as $r_0$ increases, $p_c$ increases and gradually approaches $p_{v,sat}(T_0)=3.158$ kPa, while $T_c$  decreases and approaches the boiling point, i.e. $373$ K at $101.325$ kPa. However,  pinned bubbles have lower $p_c$ and higher $T_c$ compared to unpinned ones with equal $r_0$ and $\theta_0$, or to spherical bulk bubbles with the same $R_0$ (see Appendix~\ref{appendixA}). The differences in $p_c$ and $T_c$ between pinned and unpinned bubbles are large for small $r_0$ but decrease with increasing $r_0$. In addition, as $\theta_0$ increases (i.e., the bubble becomes more flat), $p_c$ increases and $T_c$ decreases for unpinned bubbles, which is in agreement with the observed increase in cavitation  on more hydrophobic substrates~\cite{Bremond2006a,Belova2011}. In contrast, for pinned bubbles, $p_c$ decreases and $T_c$ increases with growing $\theta_0$, because flat bubbles contain less gas and therefore have lower $p_g$ and minimum $p_{ex}$ at the critical state. However, as $p_g$ at the critical state contributes little to the minimal $p_{ex}$, $p_c$ and $T_c$ only depend weakly on $\theta_0$. Similar behavior is observed under adiabatic conditions, with slightly larger magnitudes for $p_c$ and $T_c$ (see Appendix~\ref{appendixD}).

\subsection{Gas equilibration}
The results above are based on the assumption that there is no gas diffusion across the bubble surface during the expansion. This is not valid however, if the pressure or temperature perturbation is slow, such as, when the liquid is experimentally heated by a hot substrate\cite{Zhang2014a}. If there is sufficient time for equilibration to occur, the gas pressure inside the bubble ($p_g$) can be maintained at a value that depends on the gas concentration in the surrounding liquid ($c_g$) according to Henry's law, i.e., $p_{g} = c_g / H$ with $H$ the temperature-dependent Henry's solubility constant and $c_g = p_{g0} H(T_0)$ at the initial equilibrium state. If the liquid can be treated as a gas reservoir with a constant gas concentration and a uniform temperature, the unstable growth condition can be written as
\begin{equation}\label{eq:pex_henry}
p_l < p_{ex}(r,\theta,T) = p_v(T) - \frac{2\gamma(T) \sin\theta}{r} + \frac{p_{g0} H(T_0)}{H(T)}.
\end{equation}
We therefore calculated $p_c$ and $T_c$ for air surface nanobubbles initially at room temperature and pressure and subsequently evolving with the gas pressure in equilibrium according to Eq.~\ref{eq:pex_henry}. In the calculations, the empirical equation of Henry's solubility constant for the nitrogen in water ~\cite{Harvey1996} was used (see Appendix~\ref{appendixB}). The results are given in Fig.~\ref{fig:fig02} (c) and (d), and show how the gas diffusion affects $p_c$ and $T_c$. 
First, because unpinned bubbles are unstable under gas equilibration, they always have $p_c=p_{atm}$ and $T_c=T_0$ regardless of the initial size. Second, for pinned bubbles, as $r_0$ increases, $p_c$ increases and gradually approaches $p_{atm}=101.325$ kPa. At the same time, under pressure reduction $p_g=p_{g0}H(T_0)/H(T)$ can be constant rather than decreasing rapidly from $p_{g0}$. This means that $p_c$ will be smaller in magnitude than in the case of no gas diffusion, especially for bubbles with small $\theta_0$ that have high $p_{g0}$. Third, for pinned bubbles, as $r_0$ increases, $T_c$ decreases and approaches $T_0=298$ K. Furthermore, as $H$ usually decreases with increasing temperature near room temperature, for most bubbles $T_c$ will also be smaller than in the case of no gas diffusion. However, for many species $H$ goes through a minimum (\textit{e.g.} at $\sim 365$ K for nitrogen in water) and then increases with increasing temperature~\cite{Harvey1996}, so the $T_c$ of very small and flat nanobubbles may be similar to that which one would have without gas diffusion [Fig.~\ref{fig:fig02} (b) and (d)]. Also as $p_g$ contributes more to $p_{ex}$ due to the gas diffusion, the dependence of $p_c$ and $T_c$ on $\theta_0$ becomes more significant.

Overall, our theoretical analysis predicts that pinned surface nanobubbles can withstand a larger tensile stress (negative pressure) and higher temperatures (i.e., lower $p_c$ and higher $T_c$) compared to both unpinned bubbles with the same initial contact angle  $\theta_0$ and lateral radius $r_0$, and free bubbles in the bulk with the same initial radius of curvature $R_0$. Bubbles with smaller $\theta_0$ and/or $r_0$ have larger magnitudes of $p_c$ and $T_c$, even though, if there is gas diffusion, $p_c$ and $T_c$ can decrease in magnitude making the bubbles less stable. In Sec.~\ref{sec:discussion}, we will compare our theoretical analysis with previous experimental results.

\section{Simulation}

To test the validity of our theoretical analysis in the presence of more realistic fluctuations, we used many-body dissipative particle dynamics (MDPD) simulations to predict $p_c$ and $T_c$ for pinned quasi-two-dimensional (quasi-2D) bubbles in a variety of situations.

\subsection{Simulation model and method}

\begin{figure}[tb]
	\begin{center}
		\includegraphics[width=0.4\textwidth]{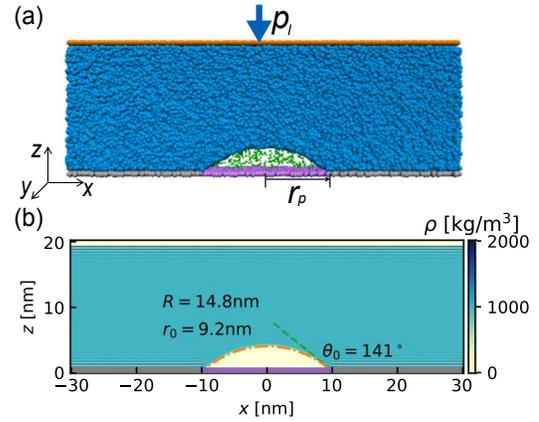}
		\caption{(a) A typical quasi-2D simulation box with a stable pinned surface nanobubble on a chemically heterogeneous surface at $T=298$ K and $p_l=0.1$ Mpa. The box dimensions are: $L_x=60$ nm, $L_y=4.25$ nm and $L_z\sim 20$ nm. The blue particles are liquid, green particles are gas, and others are part of solid walls. In the bottom wall, the purple particles form a chemically heterogeneous patch with a width of $2r_p=20$ nm. (b) Density profile of liquid particles around the pinned bubble in (a). The orange dash-dot line shows the position of the liquid/bubble interface, using a circular fit to the points at which the liquid density is half the bulk value.}
		\label{fig:fig03}
	\end{center}
\end{figure}

Figure~\ref{fig:fig03} shows a typical quasi-2D simulation box with a stable pinned surface nanobubble. The liquid and gas particles are confined between two parallel solid walls. The liquid particles are modeled as MDPD beads~\cite{Pagonabarraga2001,Arienti2011}, which interact with each other via the force $\textbf{f}_{ij} = A w_c(r_{ij}) \hat{\textbf{e}}_{ij} + B (\bar{\rho_i} + \bar{\rho_j}) w_d(r_{ij}) \hat{\textbf{e}}_{ij}$ with  $i$, $j$ particle indices. The first term represents an attractive interaction ($A<0$) and the second term a many-body repulsive interaction ($B>0$). $r_{ij}=|\textbf{r}_{ij}|$ is the distance between the two particles and $\hat{\textbf{e}}_{ij}=\textbf{r}_{ij}/r_{ij}$ is the directional unit vector. The weight functions are chosen as $w_c(r_{ij}) = 1-r_{ij}/r_c$ and $w_d(r_{ij})=1-r_{ij}/r_d$, and become zero at $r_{ij}\ge r_c$ for $w_c(r_{ij})$ and $r_{ij}\ge r_d$ for $w_d(r_{ij})$. The magnitude of the repulsion depends on the local density for each particle which is defined as $\bar{\rho_i}=\sum_{j\neq i} w_\rho(r_{ij})$ with $w_\rho(r_{ij})=15(1-r_{ij}/r_d)^2/2\pi r^3$ if $r_{ij}<r_d$ and $w_\rho(r_{ij})=0$ if $r_{ij}\ge r_d$~\cite{Pagonabarraga2001,Arienti2011,Ghoufi2011,Ghoufi2013}. 

In our simulations, all units are scaled by the bead mass $m$, the cutoff radius $r_c$, and the energy unit $e$ which is equal to the thermal energy $k_B T$ when the simulation temperature $T^*=k_BT/e = 1$ with $k_B$ the Boltzmann constant. Hence, $r^*_c=m^*=e^*=1$ . The superscript asterisk means the quantity is in reduced units. In our simulations, $r^*_d = 0.75$, $A^*=-50$ and $B^*=25$. To convert to real units, we used the following values: $r_c=8.5$ \angstrom, $m=9.0\times10^{-26}$ kg, and $e=4.1\times10^{-21}$ J, with the characteristic time $\tau=r_c\sqrt{m/e}=4$ ps. Using these parameters, the MDPD liquid at temperature $T^*=1$ and pressure $p^*=0.0015$ can approximately reproduce the thermodynamic properties such as the density and the viscosity for bulk water as well as the surface tension for the water-vapor interface at room temperature and pressure~\cite{Ghoufi2011,Ghoufi2013}.

\begin{table}[tb]
\caption{Parameters for the Lennard-Jones interaction between different particles.} % title of Table
\centering % used for centering table
\begin{tabular*}{0.4\textwidth}{l@{\extracolsep{\fill}}lll}
\hline\hline %inserts double horizontal lines
particles & $\epsilon^{\alpha\beta *}$ & $\sigma^{\alpha\beta *}$ & $r^{\alpha\beta *}_{cut}$ \\ [0.5ex] %heading
\hline % inserts single horizontal line
liquid-gas (lg) & $0.4$ & $0.7$ & $0.7\cdot (2)^{1/6}$ \\ % inserting body of the table
liquid-top (lt) & $0.35$ & $1.0$ & $2.5$ \\ 
liquid-bottom (lb) & $0.02$ or $0.4$ & $1.0$ & $2.5$ \\
gas-gas (gg) & $0.4$ & $0.4$ & $1.0$ \\ 
gas-top (gt) & $0.4$ & $0.7$ & $0.7\cdot (2)^{1/6}$ \\ 
gas-bottom (gb) & $0.4$ & $0.7$ & $0.7\cdot (2)^{1/6}$ \\ [1ex] % [1ex] adds vertical space
\hline %inserts single line
\end{tabular*}
\label{table:LJ_parameters} % is used to refer this table in the text
\end{table}

The top and bottom walls are composed of a single layer of particles placed on a simple square lattice with a lattice spacing of $0.5r_c$. The mass $m^*_g=0.5$ for gas particles and $m^*_s=1$ for solid particles. The gas and solid particles interact with each other and with liquid beads via the Lennard-Jones (LJ) potential $U^{\alpha\beta}(r)=4 \epsilon^{\alpha\beta} [(\sigma^{\alpha\beta}/{r})^{12} -(\sigma^{\alpha\beta}/{r})^6]$ truncated and shifted at $r^{\alpha\beta}_{cut}$. Here, $r$ is the distance between two particles, and $\alpha,\beta \in \{\text{liquid,gas,top-wall,bottom-wall}\}$. The LJ parameters are listed in Table~\ref{table:LJ_parameters}. The bottom wall is chemically heterogeneous in order to pin the bubble, with a hydrophobic patch ($\epsilon^{lb*}=0.02$ with the equilibrium contact angle $\sim 160^\circ$) at the center and hydrophilic regions ($\epsilon^{lb*}=0.4$ with the equilibrium contact angle $\sim 53^\circ$) elsewhere. The width of the hydrophobic patch is $2r_p$ [see Fig.~\ref{fig:fig03}(a)]. This geometry is convenient to generate ``infinite" (i.e. periodically repeated) cylindrical bubbles, whose behaviors are also governed by Eq.~\ref{eq:pex} and Eq.~\ref{eq:pex_henry} with some variations to account for the different shapes and the different Laplace pressure contributions. In principle, it is also possible to generate spherical bubbles on circular patch in larger boxes. However, the present approach is more computationally efficient, and thus yields better statistics and more accurate answers.

All simulations were carried out using the parallel molecular dynamics (MD) software package LAMMPS~\cite{Plimpton1995,Li2013}. During the simulations, the top wall was used as a barostat and was free to move as a rigid body in the horizontal and vertical directions. A constant external force along $z$ was imposed on the top wall to maintain the liquid pressure $p_l$. Each solid particle in the bottom wall was tethered to its initial position via a stiff harmonic potential with spring constant $c*=2000$. Periodic boundary conditions were imposed in the $x$ and $y$ directions. The velocity-Verlet algorithm with a time step of $0.003\tau$ (i.e., $12$ fs) was used to integrate the equation of motion, and a Nos\'{e}–Hoover thermostat with a time constant of $0.3\tau$ was used to maintain the temperature of the liquid and the bottom wall.

We simulated several surface nanobubbles of different sizes (i.e., $r_0$ and $\theta_0$). The dimensions of the simulation box were adapted to the size of the nanobubble: $L_y=4.25$ nm; $L_x=6r_p$, with $r_p=7$, $10$, or $15$ nm; and $L_z$ varied with the total number of liquid particles, which was set to $11900$, $24200$ or $121000$, respectively. All bubbles were equilibrated for $1.2$ ns at $T_0=298$ K and $p_{l0}=0.1$ MPa starting from an initial cap-shaped configuration. For the stable pinned surface nanobubble, $r_0$ was inferred  from $r_p$, and $\theta_0$ was controlled by varying the number of gas particles from $80$ to $2000$. The actual values of $r_0$ and $\theta_0$ were measured using the density profile of the liquid around the nanobubbles obtained from a $1.2$ ns simulation after equilibration [see Fig.~\ref{fig:fig03}(b) for an example].  In total, 9 surface nanobubbles with $r_0=6$--$15$ nm and $\theta=120^\circ$--$150^\circ$ were studied using pressure reduction and temperature increase protocols.

A pressure drop was obtained by instantly reducing the force imposed on the top wall, while the increase in liquid temperature was applied using a linear temperature ramp over $1$ ns, to ensure a quasi-static response of the bubble. Note that in our model, the Weeks-Chandler-Andersen (WCA) potential~\cite{Weeks1971} between the liquid and gas particles (see Table~\ref{table:LJ_parameters})  makes the gas insoluble in the liquid and therefore naturally allows us to mimic the limiting case of no mass transfer to or from the liquid. In order to take gas diffusion into account, we also performed simulations where the number of gas particles was controlled using a grand canonical Monte Carlo algorithm (GCMC)~\cite{Frenkel2001} with an ideal gas reservoir of chemical potential $\mu=k_BT_r\ln{p_r\Lambda^3/k_BT_r}$, with $p_r$ the reservoir pressure, $T_r$ the reservoir temperature and $\Lambda$ the thermal wavelength. Considering that in our model the gas is ideal [see Fig.~\ref{fig:figs04} (a) in Appendix.~\ref{appendixE}], we calculated $\mu$ using two different approaches. In the pressure reduction simulations, $\mu$ was determined by the initial gas pressure and temperature in the bubble, i.e., $p_r=p_{g0}$ and $T_r=T_0$. In the temperature increase simulations, Henry’s solubility constant cannot be easily determined in our insoluble gas model. We therefore calculated $\mu$ using $p_r=p_{g0}$ and set $T_r$ equal to the desired temperature, i.e., $H(T)=H(T_0)$. This is acceptable as we are only concerned about the consequence of gas diffusion. During the simulations, 100 GCMC exchanges were performed every 100 MD steps to achieve gas equilibration between the bubble and the surrounding liquid. All simulations were run for at least $20$ ns to ensure that the bubble was able to reach a new stable state or undergo unstable growth.

\subsection{Simulation results}

\begin{figure*}[tb]
	\begin{center}
		\includegraphics[width=0.8\textwidth]{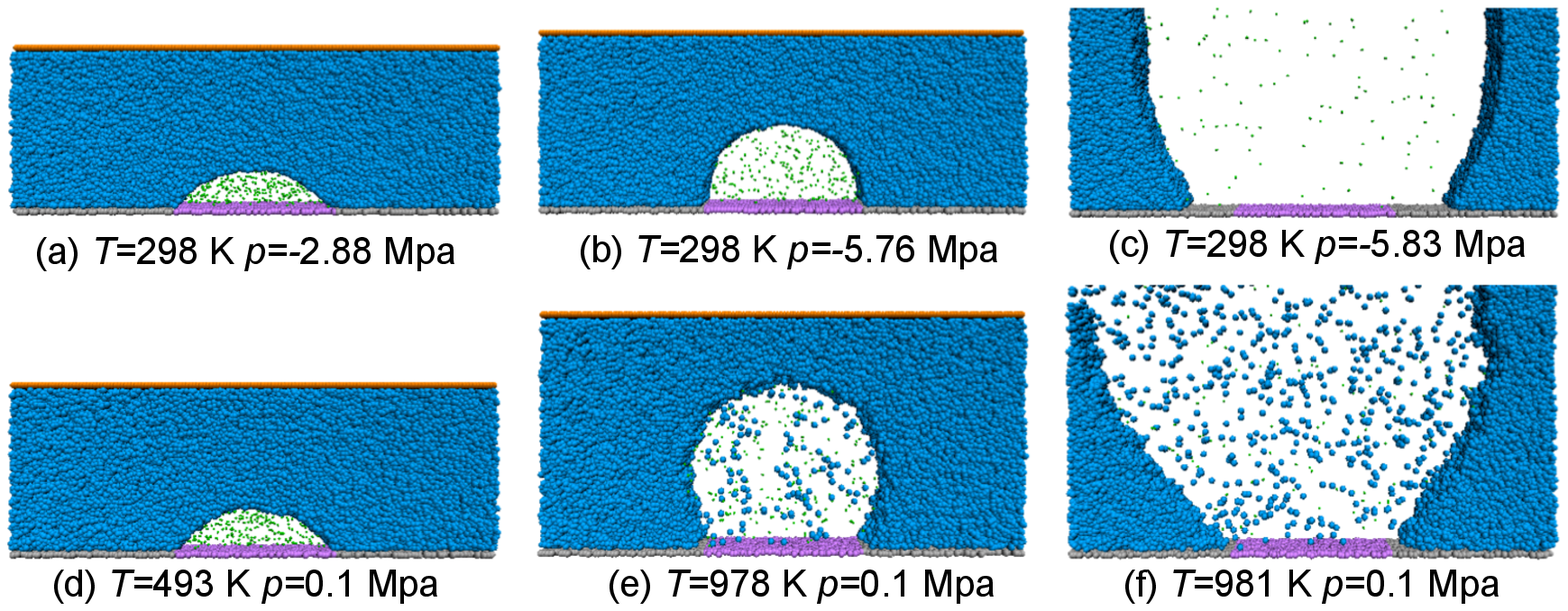}
		\caption{Snapshots of the final states obtained from the pressure reduction and temperature increase simulations, at the indicated pressures and temperatures, starting from the stable surface nanobubble in Fig.~\ref{fig:fig03} (a). The bubbles in (c) and (f) are unstable. Note that no gas diffusion across the bubble interface was allowed in these simulations.}
		\label{fig:fig04}
	\end{center}
\end{figure*}

\begin{figure}[tb]
	\begin{center}
		\includegraphics[width=0.4\textwidth]{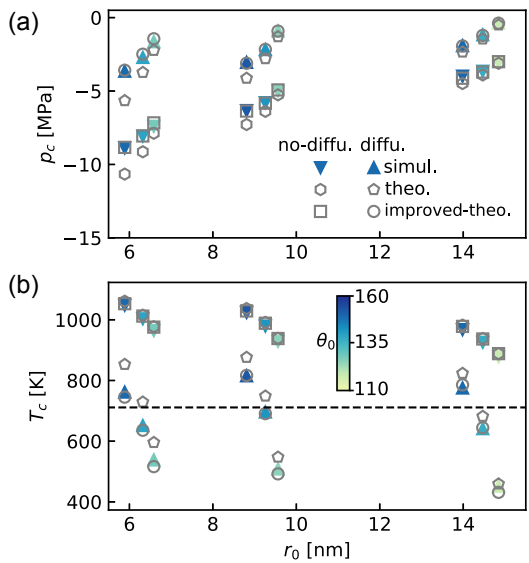}
		\caption{Critical pressures ($p_c$) and temperatures ($T_c$) for pinned surface nanobubbles with different initial values of $r_0$ and $\theta_0$. Results from the MDPD simulations and theoretical analysis (Eq.~\ref{eq:pex2D} with $m=n=1$ and Eq.~\ref{eq:pex_henry2D}) are compared. In addition, results are shown for an improved theory, in which lateral expansion of the contact line is taken into account. The horizontal dashed line in (b) corresponds to the boiling point of the MDPD liquid, which is $711$ K at $0.1$ MPa [see Fig.~\ref{fig:figs04} (b) in Appendix.~\ref{appendixE}].}
		\label{fig:fig05}
	\end{center}
\end{figure}

Figure~\ref{fig:fig04} shows some examples of final states obtained from the pressure reduction and temperature increase simulations with no GCMC exchange for the stable surface nanobubble in Fig.~\ref{fig:fig03} (a). When the pressure is reduced, the bubble expands and is able to reach a stable state at $-2.88$ Mpa and $-5.76$ Mpa. At $-5.83$ Mpa, however, the bubble exhibits unstable growth, and so we conclude that $p_c$ is somewhere between $-5.76$ and $-5.83$ Mpa. Similarly, when the temperature is increased, the bubble remains stable until $T=981$ K, indicating that $T_c$ is somewhere between $978$ and $981$ K.

The simulation results for $p_c$ and $T_c$ for surface nanobubbles with different values of $r_0$ and $\theta_0$ are plotted in Fig.~\ref{fig:fig05}. In general, these results show the same trends as those obtained from our theoretical analysis. For example, the magnitude of $p_c$ and $T_c$ decreases with increasing $r_0$ and $\theta_0$. In the absence of gas diffusion, surface nanobubbles remain stable at temperatures above the boiling point. In contrast, gas diffusion reduces the magnitude of $p_c$ and $T_c$ substantially which, for bubbles with small $\theta$, can cause $T_c$ to drop below the boiling point. 

In Fig.~\ref{fig:fig05}, we have also plotted the corresponding predictions based on the 2D version of Eq.~\ref{eq:pex} and Eq.~\ref{eq:pex_henry} under the condition of $r\equiv r_0$, i.e.,
\begin{multline}\label{eq:pex2D}
p_l < p_{ex}(r,\theta,T) = p_v(T,r,\theta) - \frac{\gamma(T) \sin\theta}{r}  \\ + p_{g0} \left( \frac{A_0}{A} \right)^n \left(\frac{T}{T_0}\right)^m,
\end{multline}
with the cross-sectional area of the 2D bubble $A=0.5r^2(\alpha-\sin\alpha)/\sin^2 \theta$ and $\alpha=2(\pi-\theta)$, and
\begin{equation}\label{eq:pex_henry2D}
p_l < p_{ex}(r,\theta,T) = p_v(T,r,\theta) - \frac{\gamma(T) \sin\theta}{r} + \frac{p_{g0} H(T_0)}{H(T)}.
\end{equation}
The thermodynamic properties of the liquid (i.e., the density, the surface tension and the saturated vapor pressure) at different temperatures were obtained by fitting the corresponding data for the MDPD liquid [see Fig.~\ref{fig:figs04} (b)-(d) in Appendix.~\ref{appendixE}]. In the simulations, the temperature of the bubble was found to always reach the liquid temperature rapidly, indicating that the whole process is almost isothermal. This is the case described by Eq.~\ref{eq:pex2D} with $m=n=1$. This expression, however, overestimates the magnitude of $p_c$ and $T_c$, especially for the small flat nanobubbles (\textit{e.g.} with $r_0\sim 6$ nm and $\theta_0\sim 150^\circ$). The reason for the discrepancy is that even in the presence of the pinning effect, the contact line exhibits slight expansion during the growth. This can be seen in the simulations (for example in Fig.~\ref{fig:fig04}), and makes $r$ increase slightly during the expansion, thereby reducing the magnitude of $p_c$ and $T_c$. To account for this, we developed an improved theory, in which the lateral expansion is described by $r=A\cos\theta + B$, where the coefficients $A$ and $B$ are determined by $r_0$ and $\theta_0$ for various stable pinned surface nanobubbles (see Appendix.~\ref{appendixF}). As can be seen in Fig.~\ref{fig:fig05}, the improved theory is much better at predicting $p_c$ and $T_c$, indicating that the discrepancies from the original theory with $r\equiv r_0$ are mainly due to the lateral expansion. The improved classical theory can predict our simulaton results for the stability of pinned surface nanobubbles as small as several nanometers.

\section{Discussion}\label{sec:discussion}

So far, we have presented our results for the pressure and temperature thresholds ($p_c$ and $T_c$) beyond which unstable growth occurs for surface nanobubbles when the pressure is reduced or the temperature is increased. In this section, we compare our results with other studies.

Our results for pinned surface nanobubbles under pressure reduction are consistent with those recently reported by Dockar \textit{et al.}~\cite{Dockar2019}. From Eq.~\ref{eq:pex} and Eq.~\ref{eq:pex2D}, by computing $p_c=\min(p_{ex})$ at $\partial p_{ex} / \partial \theta = 0$, we can reproduce the equations for the corrected cavitation threshold derived by Dockar \textit{et al.}. In their theoretical calculations they assumed that the maximum gas-side contact angle that a surface nanobubble can sustain is $90^\circ$. This assumption was perfectly reasonable, because in their simulations the contact line would unpin at $90^\circ$ (the equilibrium contact angle on the hydrophilic region), meaning that the gas-side contact angle could never exceed $90^\circ$ during the expansion. However, if the patch is more hydrophilic, as in our simulations, the pinned bubble will not reach the critical point until the gas-side contact angle is larger than $90^\circ$ (see Fig.~\ref{fig:fig04}). Our analytical theory is able to describe this case as well, without any constraints on $\theta$ (see Fig.~\ref{fig:fig01} for example). 

Our results also show that the movement of the contact line can significantly affect $p_c$ and $T_c$. For example, the "stick-jump” behavior in Dockar \textit{et al.}'s simulations decreases the magnitudes of $p_c$ and $T_c$ compared to that expected for ideally pinned bubbles. Another example is the slight lateral expansion of the contact line observed in our simulations during growth, which also decreases the magnitudes of $p_c$ and $T_c$ (see Fig.~\ref{fig:fig05}). The actual behavior of the contact line in an experimental system will strongly depend on the physical and chemical properties of the substrate and could be very complex~\cite{Snoeijer2013,Liu2014a}. The theoretical results in the case of ideal pinning (i.e., $r\equiv r_0$) describe maximally stable surface nanobubbles, i.e., they predict extrema for $p_c$ and $T_c$. In order to obtain more precise predictions from classical theory, the effect of the contact line dynamics needs be taken into account, such as in our modified theory that includes the effect of the lateral expansion.

Our theoretical calculations (both original and modified) predict that surface nanobubbles with $r_0>25$ nm should undergo unstable growth at $-6$ Mpa, but an earlier experiment~\cite{Borkent2007} reported that surface nanobubbles with $r_0=30$--$150$ nm remain stable up to $-6$ Mpa [see Fig.~\ref{fig:fig02} (a) and (c)]. This disagreement could be due to the specific protocol followed in the experiment where the pressure perturbation was generated by a shock wave with both compression and expansion phases. As pointed out by the authors, the original nanobubbles may be removed during the initial rapid compression phase and different bubbles could form under the following negative pressure~\cite{Borkent2007}. Dockar \textit{et al.} also noticed this discrepancy and hypothesized a possible mechanism by which the bubbles detaching from the surface during unstable growth could form more stable bulk nanobubbles with a reduced radius of curvature~\cite{Dockar2019}. It is also possible that contamination~\cite{Ducker2009}, surface charges~\cite{Yurchenko2016} or collective effects~\cite{Bremond2006a} could be responsible for the abnormal stability reported in these experiments.

Finally, the upper critical temperature for stability of surface nanobubbles has also been investigated experimentally~\cite{Zhang2014a}. We have plotted this data as crosses in Fig.~\ref{fig:fig02} (b) and (d). If there is no gas diffusion, all pinned surface nanobubbles are predicted to be stable below the boiling temperature, and the classical theory is able to explain the experimental results. However, gas diffusion should be non-negligible in these experiments because the liquid was slowly heated by a hot substrate. If the gas diffusion was sufficiently fast to allow for equilibration between the bubble and the water, as we considered above, then only small flat nanobubbles should be stable. A possible reason for this discrepancy is that diffusion of gas also occurs from the water into the surrounding environment, thus reducing the gas concentration in the liquid and moving $T_c$ back towards the boiling point.

\section{Summary}
In this work, we investigated the stability of pinned surface nanobubbles exposed to a reduction in pressure or an increase in temperature. The pressure and temperature thresholds ($p_c$ and $T_c$) at which unstable growth occurs for surface nanobubbles in water were first determined using the expansion-collapse criterion from the classical Rayleigh-Plesset theory (Eq.~\ref{eq:pex} and Eq.~\ref{eq:pex_henry}). Due to contact line pinning, surface nanobubbles grow with a nearly constant lateral radius, and consequently a reduced radius of curvature, resulting in a lower $p_c$ and higher $T_c$ compared to unpinned bubbles with the same initial contact angle and lateral radius ($\theta_0$ and $r_0$) or bulk bubbles with the same initial radius of curvature ($R_0$). $p_c$ and $T_c$ are dependent on both $\theta_0$ and $r_0$, and decreasing $\theta_0$ and/or $r_0$ increases the magnitude of $p_c$ and $T_c$, making the bubbles more stable. On the other hand, gas diffusion between the bubble and the liquid can decrease the magnitude of $p_c$ and $T_c$. The values of $p_c$ and $T_c$ obtained from MDPD simulations of quasi-2D surface nanobubbles are consistent with our theoretical predictions, provided that we account for the lateral expansion of the contact line that occurs in our simulations during bubble growth. This validates our classical thermodynamic description for pinned surface nanobubbles as small as several nanometers. Even though our predictions for $p_c$ and $T_c$ are inconsistent with some experimental results, we would like to point out that these discrepancies could be attributed to the high uncertainties in bubble dynamics and gas diffusion inherent in current experimental techniques. We hope that our work will inspire further theoretical and experimental studies that will ultimately yield a more comprehensive understanding of surface nanobubble stability.

%--------------------------------------------------------------------------------------------
%============================================================================================
% Acknowledgements
%============================================================================================
%--------------------------------------------------------------------------------------------

\begin{acknowledgements}
This work was supported by the Australian Research Council under Grant CE170100026. Computational resources were provided by the University of Sydney HPC service, and the Pawsey Supercomputing Centre with funding from the Australian Government and the Government of Western Australia.
\end{acknowledgements}

\bibliography{text}

\clearpage
\onecolumngrid

%--------------------------------------------------------------------------------------------
%============================================================================================
% Appendix
%============================================================================================
%--------------------------------------------------------------------------------------------
\appendix

%============================================================================================
% section
%============================================================================================
\section{Unstable growth condition for free spherical bubbles}\label{appendixA}
Similarly to what shown in the main text, it is easy to derive the unstable growth condition for free spherical bubbles with initial radius $R_0$ and initial gas pressure $p_{g0}=p_{atm}+ 2\gamma(T_0)/R_0 - p_v(T_0)$, i.e.
\begin{equation}\label{eq:pex_bulk}
p_l < p_{ex}(R,T) = p_v(T) - \frac{2\gamma(T)}{R} + p_{g0} \left( \frac{3V^B_0}{4\pi R^3} \right)^n \left(\frac{T}{T_0}\right)^m.
\end{equation}
with $V^B_0$ the initial volume of the bulk bubble. The minimum liquid pressure that satisfies Eq.~\ref{eq:pex_bulk} is often called the Blake threshold for bubbles under pressure reduction (i.e., $T=T_0$)~\cite{Blake1949,Brennen1995} and can be obtained at $\partial p_{ex} / \partial R = 0$, i.e.,
\begin{equation}\label{eq:Blake_threshold}
p_B = \min(p_{ex}) = p_v - 2\gamma \left(1-\frac{1}{3n}\right) \left(\frac{\gamma}{3nR_0^{3n}p_{g0}}\right)^{1/(3n-1)}.
\end{equation}

The unstable growth condition for surface nanobubbles can also be written with respect to the radius of curvature $R$, and the initial bubble volume $V_0=V_0^B g(\theta_0) / 4$ with $V^B_0$ the volume of a bulk spherical bubble of the same $R_0$. Thus, Eq.~\ref{eq:pex} can be rewritten as
\begin{equation}\label{eq:pex_R}
p_l < p_{ex}(R,\theta,T) = p_v(T) - \frac{2\gamma(T)}{R}  + p_{g0} \left( \frac{3V_0^B g(\theta_0)}{4\pi R^3 g(\theta)} \right)^n \left(\frac{T}{T_0}\right)^m.
\end{equation}
If the surface nanobubble evolves with a constant contact angle (i.e., $\theta \equiv \theta_0$), Eq.~\ref{eq:pex_R} will become exactly the same as Eq.~\ref{eq:pex_bulk} because $g(\theta) \equiv g(\theta_0)$ and will lead to the same Blake threshold as Eq.~\ref{eq:Blake_threshold}. Therefore, an unpinned surface nanobubble should have the same $p_c$ and $T_c$ as that of a free bubble with the same $R_0$~\cite{Bremond2006a}.

%============================================================================================
% section
%============================================================================================
\setcounter{figure}{0} 
\counterwithin{figure}{section}
\section{Thermodynamic properties of water and air}\label{appendixB}

The thermodynamic properties of water and Henry’s solubility constant for nitrogen in water were calculated using empirical equations~\cite{Vargaftik1983,Wagner1993,Harvey1996}. The critical temperature $T_c=647.15$ K, the critical saturated vapor pressure $p_{v,c}=22.064\times 10^6$ Pa and the critical liquid density $\rho_c=322$ kg/m$^3$.
\begin{itemize}
    \item The saturated vapor pressure ($p_{v,sat}$ [Pa]) is given by
\begin{equation}
\ln\left(\frac{p_{v,sat}}{p_{v,c}}\right) =
\frac{T_c}{T} \left[ a_1\tau + a_2\tau^{1.5} + a_3\tau^3 + a_4\tau^{3.5} + a_5\tau^{4} + + a_6\tau^{7.5} \right]
\end{equation}
    with $a_1 = -7.85951783$, $a_2 =  1.84408259$, $a_3 = -11.7866497$, $a_4 =  22.6807411$, $a_5 = -15.9618719$, and $a_6 =  1.80122502$.
    \item The liquid density ($\rho_l$ [kg/m$^3$]) is given by
\begin{equation}
\frac{\rho_l}{\rho_c} =
1 + b_1\tau^{1/3} + b_2\tau^{2/3} + b_3\tau^{5/3} + b_4\tau^{16/3} + b_5\tau^{43/3} + b_6\tau^{110/3}
\end{equation}
    with $b_1 = 1.99274064$, $b_2 =  1.09965342$, $b_3 = -0.510839303$, $b_4 =  -1.75493479$, $b_5 = -45.5170352$, and $b_6 =   -6.74694450\times10^5$. 
    \item The surface tension ($\gamma$ [N/m]) is given by
\begin{equation}\label{eq:vpaor_press}
\gamma = B\left[\frac{T_c-T}{T_c}\right]^{\mu} \left[1+b\left(\frac{T_c-T}{T_c}\right)\right],
\end{equation}
    with $B=0.2358$, $\mu=1.256$, $b=-0.625$.
    \item Henry's solubility constant ($H$ [mol/m$^3$/MPa]) for nitrogen in water is given by
\begin{equation}\label{eq:henry_constant}
\ln \frac{M}{H\rho_l} = \frac{A}{T/T_c} +B\frac{(1-T/T_c)^{0.355}}{T/T_c} 
+ C \exp(1-T/T_c)(T/T_c)^{-0.41} + \ln p_{v,sat}
\end{equation}
    with $B=-11.6184$, $B=4.9266$, $C=13.3445$.
\end{itemize}

%============================================================================================
% section
%============================================================================================
\setcounter{figure}{0} 
\counterwithin{figure}{section}
\section{Change of excess pressure at different temperatures}\label{appendixC}
\begin{figure*}[ht]
	\begin{center}
		\includegraphics[width=0.8\textwidth]{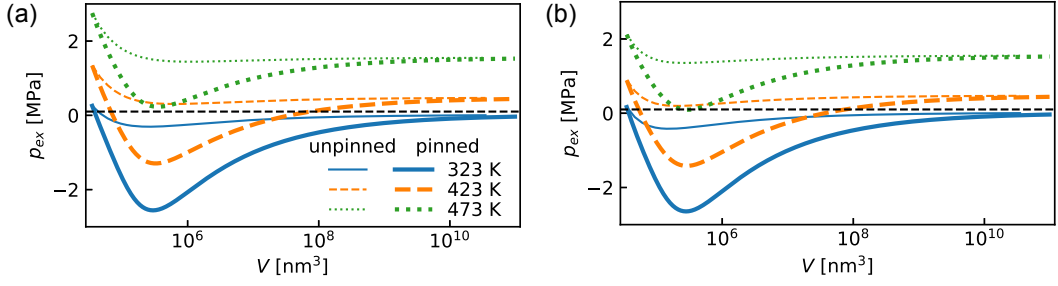}
		\caption{Change in excess pressure $p_{ex}$ (see Eq.~\ref{eq:pex}) for growing surface nanobubbles with $r_0=50$ nm and $\theta_0=160^{\circ}$, initially at room temperature and pressure, and then undergoing isothermal (a) or adiabatic (b) evolution at three different temperatures. The horizontal dashed line corresponds to $p_{l}=p_{atm}=101.325$ kPa.}
		\label{fig:figs02}
	\end{center}
\end{figure*}

%============================================================================================
% section
%============================================================================================
\setcounter{figure}{0} 
\counterwithin{figure}{section}
\section{Pressure and temperature thresholds for bubbles under adiabatic condition}\label{appendixD}
\begin{figure*}[ht]
	\begin{center}
		\includegraphics[width=0.8\textwidth]{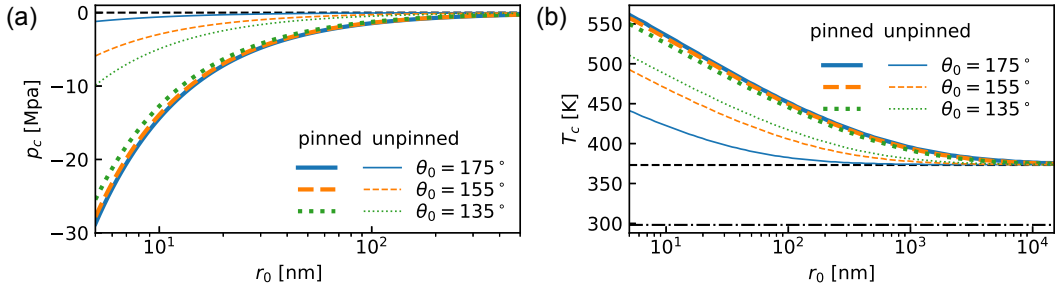}
		\caption{$p_c$ (a) and $T_c$ (b) as a function of $r_0$ (see Eq.~\ref{eq:pex}) at $\theta_0=135^\circ$, $155^\circ$ and $175^\circ$ for surface nanobubbles  initially at room temperature and pressure, and then undergoing adiabatic evolution. The horizontal dashed lines correspond to $p_{v,sat}=3.158$ kPa in (a), and the boiling point $373$ K and room temperature $298$ K in (b).}
		\label{fig:figs03}
	\end{center}
\end{figure*}

%============================================================================================
% section
%============================================================================================
\clearpage
\setcounter{figure}{0} 
\counterwithin{figure}{section}
\section{Thermodynamic properties of gas and liquid in the MDPD simulation model}\label{appendixE}
\begin{figure*}[ht]
	\begin{center}
		\includegraphics[width=0.8\textwidth]{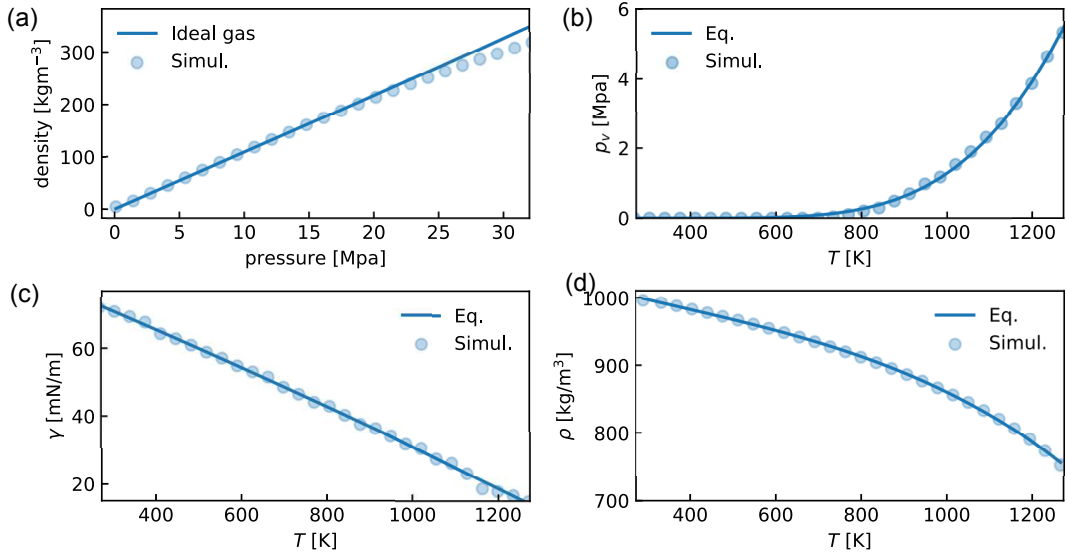}
		\caption{ (a) Gas density at different pressures and at $298$ K. The solid line is the equation-of-state for the ideal gas. The data were obtained from a simulation with $1000$ gas particles in a cubic box. (b-d) Thermodynamic properties of the MDPD liquid. (b) Saturated vapor pressure at different temperatures. The boiling point at $0.1$ Mpa is $711$ K. The fitting equation is $\log_{10}p_v = A-(B/(C+T))$ with $A = 9.41847$, $B = 3833.13$ and $C = 157.658$.  (c) Surface tension for the liquid-vapor interface at different temperatures. At $298$ K, the surface tension is $71$ mN/m, which is very close to the value of water at the same temprature. The fitting equation is $\gamma = aT^2+bT+c$ with $a = -3.97652\times10^{-9}$, $b = -5.22706\times10^{-5}$  and $c = 8.70742\times10^{-2}$. (d) Density at different temperatures at $0.1$ Mpa. At $298$ K, the density is $997$ kg/m$^3$.   The fitting equation is $\rho = aT^3+bT^2+cT+d$ with $a=-1.42055\times10^{-7}$, $b = 1.63578\times10^{-4}$, $c= -2.11321\times10^{-1}$ and $d= 1.04993\times10^{9}$. The data for saturated vapor pressure and surface tension were obtained from a simulation with one liquid slab consisting of $1000$ MDPD particles in a rectangular simulation box with edges $L_x=L_y=8.5$ nm and $L_y=17$ nm. The data for the density were obtained from a simulation with $1000$ MDPD particles in a cubic box. }
		\label{fig:figs04}
	\end{center}
\end{figure*}

%============================================================================================
% section
%============================================================================================
\setcounter{figure}{0} 
\counterwithin{figure}{section}
\setcounter{table}{0} 
\counterwithin{table}{section}
\section{Relation between lateral radii and contact angles for pinned surface nanobubbles}\label{appendixF}
\begin{figure}[ht]
	\begin{center}
		\includegraphics[width=0.4\textwidth]{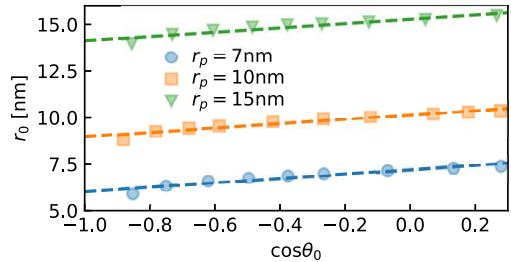}
		\caption{Dependence of $r_0$ on $\cos\theta_0$ for stable surface nanobubbles when $r_p=7$, $10$ and $15$ nm. The dashed lines show the linear fits listed in Table.~\ref{table:coefficients}}
		\label{fig:figs01}
	\end{center}
\end{figure}

In our simulations, we found that the lateral radius ($r$) is approximately a linear function of the cosine of the liquid-side contact angle ($\theta$) for stable pinned surface nanobubbles on substrates with a hydrophobic patch of width $2r_p$ (see Fig.~\ref{fig:figs01}). We assume that when the surface nanobubble expands quasi-statically, the lateral radius obeys the same linear relationship between $r$ and $\cos\theta$, i.e., $r=A\cos\theta + B$. The values of $A$ and $B$ at several different values of $r_p$ are given in Table.~\ref{table:coefficients}.

\begin{table}[ht]
\caption{Coefficients for the linear fit $r=A\cos\theta + B$ at the indicated values of $r_p$.} % title of Table
\centering % used for centering table
\begin{tabular*}{0.8\textwidth}{l@{\extracolsep{\fill}}lll}
\hline\hline %inserts double horizontal lines
$r_p$ [nm] & $A$ [nm] & $B$ [nm] \\ [0.5ex] %heading
\hline % inserts single horizontal line
$7$ & $1.480$ & $7.376$  \\ % inserting body of the table
$10$ & $1.488$ & $10.345$  \\
$15$ & $1.437$ & $15.433$  \\ [1ex] % [1ex] adds vertical space
\hline %inserts single line
\end{tabular*}
\label{table:coefficients} % is used to refer this table in the text
\end{table}

\end{document}